\documentclass[aps,prl,groupedaddress,fleqn,twocolumn]{revtex4}
\pdfoutput=1
\usepackage{graphicx}
\usepackage{amsmath}
\usepackage{gensymb}
\begin{document}

\title{Dissipation and energy gap}
\author{K. Trachenko$^{1}$}
\address{$^1$ School of Physics and Astronomy, Queen Mary University of London, Mile End Road, London, E1 4NS, UK}

\begin{abstract}
The effect of anharmonicity (coupling) in the field theory generally result in dissipation of plane waves. It has been appreciated that anharmonicity and ensuing dissipation of plane waves can be accompanied by the emergence of the gapped momentum state. Here, we show that the same effect can lead to a gapped energy state and a dispersion relation where the frequency (energy) gap emerges explicitly. We discuss several notable properties of gapped energy and momentum states and connections between them.
\end{abstract}

%\pacs{61.43.Fs, 64.70.Pf, 61.20.Lc}

\maketitle

The subject of quantum field theory is rooted in a harmonic paradigm \cite{lzee} represented by the Lagrangian as

\begin{equation}
L=\frac{1}{2}\left(\sum\limits_a\mu\dot{q}_a^2-\sum\limits_{a,b}k_{ab}q_a\,q_b-\sum\limits_{a,b,c}g_{abc}q_aq_bq_c-\dots \right)
\label{lzee}
\end{equation}

\noindent where $q$ are displacements corresponding to field variables in the field theory.

The first two terms correspond to harmonic theory and plane wave solutions. The remaining terms are due to anharmonic effects resulting in scattering, interaction and production of particles. Perturbation theory can be used when the anharmonic terms are small. A problem arises when anharmonic effects are large and coupling is strong as in many important cases. For example, the ability of liquids to flow is represented by large anharmonic terms. As a result, a first-principles theory of liquids involves a large number of coupled anharmonic oscillators and becomes exponentially complex \cite{ropp}.

Despite the absence of a small parameter in liquids, progress in liquid physics is made by introducing a property characterising the anharmonic potential, the liquid relaxation time $\tau$. $\tau$ is related to liquid viscosity by Maxwell relation and quantifies the time between molecular jumps in the liquid  \cite{frenkel} due to anharmonicity. As recently reviewed \cite{ropp,physrep}, this process leads to wave dissipation with decay time set by $\tau$ and, importantly, modifies the transverse phonon dispersion relation which acquires a gap in $k$ (momentum) space, resulting in a gapped momentum state (GMS).

We note that the effect of any anharmonicity is generally to introduce dissipation of plane waves given by the first two terms in Eq. (\ref{lzee}) \cite{ropp}. Indeed, the plane waves are eigenstates of the harmonic crystal and do not decay. However, they are not eigenstates of the total Lagrangian (these eigenstates are generally unknown). Therefore, plane waves decay, or dissipate, in a theory given by (\ref{lzee}). This is a generic effect unrelated to the type or strength of the anharmonic terms (coupling). The above strong anharmonicity in the liquid state is a subset of this effect and is related to a multi-valley potential landscape giving rise to molecular jumps \cite{ropp}.

Here, we show that the strong anharmonicity and ensuing dissipation can be accompanied by a {\it gapped energy state} emerging at a classical level, with a dispersion relation with the energy gap emerging explicitly. In this picture, gaps in momentum and energy emerge on equal footing as an accompanying effect of dissipation due to strong anharmonicity of interaction in Eq. (\ref{lzee}). We discuss important connections between gapped energy and momentum states.

We note that our discussion does not involve dissipation or change of the total energy of the system related to \eqref{lzee}. The total energy, either constant in an isolated system or fluctuating if connected to a thermal reservoir, does not decay. The decaying object are the plane waves because they are not eigenstates in a coupled system as discussed above, and it is this dissipation that we discuss here. This has important consequences for the quantum theory because, for example, canonical quantization involves non-decaying plane waves.

We start by recalling the origin of GMS as this is important for the subsequent discussion of the energy gap. The GMS emerges in several areas of physics, including liquids, supercritical fluids, plasma, Sine-Gordon model, relativistic hydrodynamics and holographic models \cite{physrep}. In liquid physics, the GMS emerges as a solution of the equation for the shear velocity field $v$ \cite{ropp,physrep}:

\begin{equation}
c^2\frac{\partial^2v}{\partial x^2}=\frac{\partial^2v}{\partial t^2}+\frac{1}{\tau}\frac{\partial v}{\partial t}
\label{gms1}
\end{equation}

\noindent where $c$ is the transverse speed of sound in the solid.

The last term in Eq. (\ref{gms1}) results in dissipation and represents the reduction of the problem of liquid theory by introducing liquid relaxation time \cite{ropp}. Eq. (\ref{gms1}) is related to strong anharmonicity in Eq. \eqref{lzee} where the multi-valley potential enables large particle jumps with liquid relaxation time independent of $k$. Weak anharmonicity in Eq. \eqref{lzee}, corresponding to an anharmonic solid without particle jumps, also results in wave dissipation, however in this case the $k$-gap is small and unimportant because the phonon lifetime increases at small $k$. Eq. \eqref{gms1} therefore represents a subset of dynamical equations describing dissipation of plane waves; the more general case is given by Eq. \eqref{lzee}.

Eq. (\ref{gms1}) is a simplified form of the equation written by Frenkel \cite{ropp} who used the Maxwell idea that liquids are capable of both elastic and viscous response and are viscoelastic \cite{maxwell}. Frenkel modified \cite{frenkel} the Navier-Stokes equation by treating viscosity as an operator which includes an elastic response according to the Maxwell proposal, resulting in Eq. (\ref{gms1}) \cite{ropp}. The same equation follows from modifying the elastic constitutive relation by generalizing the shear modulus to include the viscous response \cite{pre}.

Eq. \eqref{gms1} has the form of the ``telegraph'' or ``Telegraphist's'' equation, the term attributed to Poincare \cite{kosh}. It was derived by Heaviside \cite{heaviside} and earlier by Kirchhoff in 1857 \cite{math2}. It is a surprising fact that despite the long history of the telegraph equation, its different dispersion relations and their properties are not commonly discussed \cite{math2,math1,heat,masoliver}.

We discuss the dispersion relation for plane waves rather than for full solutions of \eqref{lzee}, as (a) the full solutions are generally unknown and (b) plane-wave excitations are of central importance in field theory including quantization, field operators and their use in treating interactions and scattering. Seeking the solution of (\ref{gms1}) as $v\propto\exp\left(i(kx-\omega t)\right)$ gives

\begin{equation}
\omega^2+\omega\frac{i}{\tau}-c^2k^2=0
\label{gms2}
\end{equation}

The full solution of (\ref{gms1}) depends on initial and boundary conditions. Considering real $k$ and complex $\omega$ gives decay in time. A model example corresponding to these boundary conditions is a propagating wave set up in an elastic rod which is then immersed in a viscous liquid. The wave does not vary with distance but decays in time until the wave oscillations cease. Apart from the model example, this case corresponds to phonon decay in liquids \cite{pre,prl}. In this case, the GMS emerges {\it explicitly}: the solution of (\ref{gms2}) is

\begin{equation}
\omega=-\frac{i}{2\tau}\pm\sqrt{c^2k^2-\frac{1}{4\tau^2}}
\label{gms3}
\end{equation}

\noindent giving time decay and dissipation of transverse waves as

\begin{equation}
v\propto\exp\left(-\frac{t}{2\tau}\right)\exp\left(i(kx-\omega_r t)\right)
\label{gms4}
\end{equation}

\noindent where

\begin{equation}
\omega_r=\sqrt{c^2k^2-\frac{1}{4\tau^2}}
\label{gms5}
\end{equation}

\noindent is real if $k$ is larger than the threshold value $k_g$ given by

\begin{equation}
k_g=\frac{1}{2c\tau}
\label{gms6}
\end{equation}

Eqs. \eqref{gms5} and \eqref{gms6} describe the gapped momentum state. It is observed in high-temperature liquids and supercritical fluids, and $k_g$ is found to decrease with $\tau$ and increase with temperature \cite{prl} in accordance with \eqref{gms6}.

The dispersion relation (DR) (\ref{gms5}) showing the GMS is shown in Fig. 1. We recognize that GMS is not commonly discussed despite the long history of the telegraph equation \cite{kosh,heaviside,math1,math2,heat,masoliver}, as compared to two other DRs: gapless phonon or photon DR and the DR of a massive particle.

In (\ref{gms4}), the decay time and decay rate are $2\tau$ and $\Gamma=\frac{1}{2\tau}$. If, as is often assumed, the crossover between propagating and non-propagating modes is given by the equality between the decay time and inverse frequency, $\omega\Gamma=1$, the propagating waves conditionally correspond to $k>\frac{1}{\sqrt{2}}\frac{1}{c\tau}$ from (\ref{gms5}), or $k>k_g\sqrt{2}$ \cite{physrep}. According to \eqref{gms5}, this gives propagating modes above the frequency

\begin{equation}
\omega=\frac{1}{2\tau}
\label{omprop}
\end{equation}

Similar result for propagating shear waves in liquids was obtained by Frenkel \cite{frenkel} by analysing the complex shear modulus derived from Maxwell's viscoelastic relation, which is the basis for \eqref{gms1}. We will later see that the frequency gap arises explicitly in the dispersion relation in a different model of wave dissipation.

%We note that this is an approximation because the condition $\omega\Gamma=1$ is not rigorous. %The case of the frequency gap discussed below will be free from this approximation.

Microscopically, the gap in $k$-space can be related to a finite propagation length of waves in a liquid: if $\tau$ is the time during which stress (e.g. shear stress) relaxes, $c\tau$ gives the shear wave propagation length. Therefore, the condition $k>k_g=\frac{1}{2c\tau}$ approximately corresponds to propagating waves with wavelengths shorter than the propagation length. %In the above example of the rod immersed in the viscous liquid, the $k$-gap corresponds to the absence of wavelengths larger than $c\tau$ because the wave disappears after time $\tau$.

\begin{figure}
\begin{center}
{\scalebox{0.37}{\includegraphics{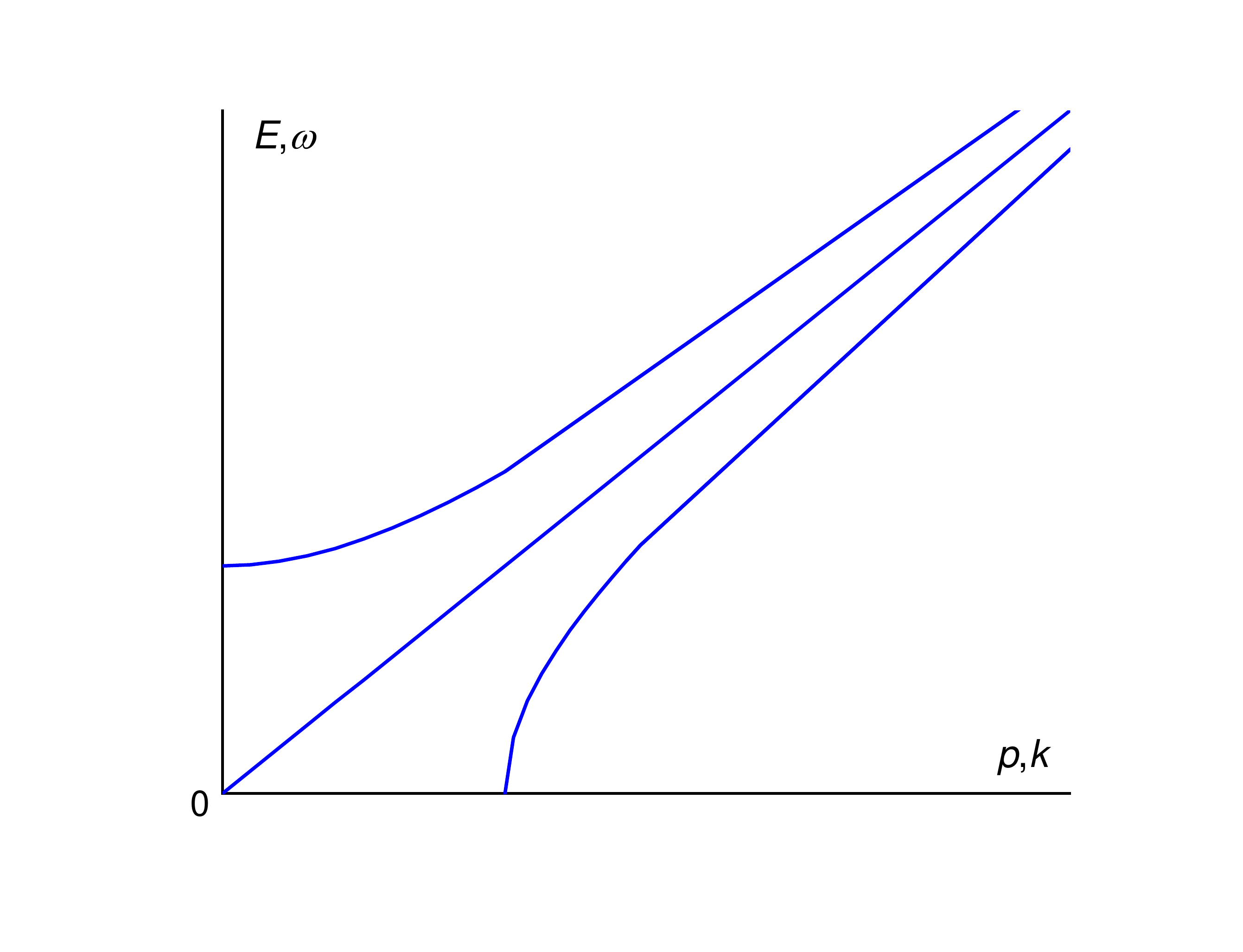}}}
\end{center}
\caption{Three different dispersion relations: dependencies of energy $E$ or frequency $\omega$ on momentum $p$ or wavevector $k$. The middle straight line shows gapless dispersion relation for a massless particle (photon) or a phonon in solids. The bottom curve shows the dispersion relation of the gapped momentum state (\ref{gms5}). The top curve shows the dispersion relation of the gapped energy state \eqref{ges7}. Schematic illustration.}
\label{disp}
\end{figure}

Our last observation regarding GMS which will become useful later is that common discussions consider complex $k$ and real $\omega$, corresponding to decay in space (see, e.g., Refs. \cite{math2,skin}). A model example of this is different from the first case and corresponds the boundary condition where a wave is induced from one end of an elastic rod immersed in a viscous liquid. Using complex $k$ and real $\omega$ in Eq. \eqref{gms2} and accompanying decay in space corresponds to the propagation of electromagnetic waves in a conductor and gives the skin effect (see, e.g., Ref. \cite{skin}). In this case, the wave persists at all times but decays with distance. Notably, the DR in this case is different from \eqref{gms5}. (This may come as surprising, given that it follows from the same Eq. \eqref{gms1} or \eqref{gms2}, yet it was noted that different dispersion relations can originate from the same equation depending on boundary conditions \cite{different} and experimental setup \cite{different1}.) The GMS due to dissipation emerges in this case implicitly. Indeed, the solution of (\ref{gms2}) is $k=k_1+ik_2$, where

\begin{equation}
k_1=\frac{\omega}{c\sqrt{2}}\left(\sqrt{1+\left(\frac{1}{\omega\tau}\right)^2}+1\right)^\frac{1}{2}
\label{gms7}
\end{equation}
\noindent and

\begin{equation}
k_2=\frac{\omega}{c\sqrt{2}}\left(\sqrt{1+\left(\frac{1}{\omega\tau}\right)^2}-1\right)^\frac{1}{2}
\label{gms8}
\end{equation}

\noindent resulting in $v\propto e^{-k_2x}e^{i(k_1x-\omega t)}$.

The DR \eqref{gms7} is gapless, which is easier to see if its written as $\omega=\frac{2c^2k^2}{\sqrt{4c^2k^2+\frac{1}{\tau^2}}}$. In the propagating regime $\omega\tau\gg 1$, $k_1=\frac{\omega}{c}$ and $k_2=\frac{1}{2c\tau}$, giving space-decaying field as

\begin{equation}
v\propto e^{-\frac{x}{2c\tau}}e^{i(k_1x-\omega t)}
\label{gms10}
\end{equation}

For distances smaller than $c\tau$, the wave propagation in Eq. \eqref{gms10} proceeds as in the absence of dissipation. However, the $k$-gap emerges implicitly because (\ref{gms10}) implies the decay range $2c\tau$ and hence no propagating plane waves with $k<\frac{1}{2c\tau}$, or below $k_g$ in \eqref{gms6}.

We now discuss how the energy gap emerges in the DR explicitly. We recall that Eq. (\ref{gms1}) describes the dynamics of liquids in the Maxwell-Frenkel viscoelastic theory. The same equation can be derived in a simple model which we will later use to discuss the energy gap. The model is an elastic rod undergoing vibrations (e.g., longitudinal vibrations) in the presence of a dissipative force. Applying the balance of forces to the element of the rod with length $\Delta x$ gives

\begin{equation}
\rho S\Delta x\frac{\partial^2u}{\partial t^2}=SL(\epsilon(x+\Delta x)-\epsilon(x))+F_d
\label{ges1}
\end{equation}

\noindent where $u$ is displacement, $\rho$ is density, $S$ is cross-section area, $L$ is the modulus of elasticity, $\epsilon=\frac{\partial u}{\partial x}$ is strain and $F_d$ is a dissipative force.

If dissipation is due to, for example, motion in a viscous liquid, $F_d\propto-\Delta x\frac{\partial u}{\partial t}$. Then, Eq. (\ref{ges1}) becomes

\begin{equation}
\frac{\partial^2u}{\partial t^2}=c^2\frac{\partial^2u}{\partial x^2}-\frac{1}{\tau}\frac{\partial u}{\partial t}
\label{ges11}
\end{equation}

\noindent where the factor $\frac{1}{\tau}$ includes the proportionality coefficient between $F_d$ and $\frac{\partial u}{\partial t}$ and where we used $L=\rho c^2$.

Eq. (\ref{ges11}) is the same as our starting equation (\ref{gms1}) which gives the GMS.

We now observe that although invoking the usual friction force $F_d\propto-\frac{\partial u}{\partial t}$ is a common way to discuss dissipation in the field of mathematical physics and partial differential equations \cite{math1,math2}, it is not the only possibility to obtain dissipation. Another way of achieving decay and dissipation is to consider $F_d$ due to an external source or field which is proportional to the space rather than time derivative of $u$, or strain: $F_d\propto\frac{\partial u}{\partial x}$, where the axis $x$ is along the displacement $u$. Then, Eq. (\ref{ges1}) becomes

\begin{equation}
\frac{\partial^2u}{\partial t^2}=c^2\frac{\partial^2u}{\partial x^2}+\frac{c}{\tau}\frac{\partial u}{\partial x}
\label{ges2}
\end{equation}

\noindent where the factor $\frac{c}{\tau}$ includes the proportionality coefficient between $F_d$ and $\frac{\partial u}{\partial x}$ and we keep the same parameter $\tau$ for simplicity.

The spectrum of \eqref{ges2} has the energy gap. This is most easily seen by using the standard transformation eliminating the first derivatives in hyperbolic equations \cite{math2,math1}. Using $u=\phi\exp\left(-\frac{x}{2c\tau}\right)$ in (\ref{ges2}) gives the Klein-Gordon equation

\begin{equation}
c^2\frac{\partial^2\phi}{\partial x^2}=\frac{\partial^2\phi}{\partial t^2}+\frac{\phi}{4\tau^2}
\label{t2}
\end{equation}

\noindent with the spectrum

\begin{equation}
\omega^2=c^2k^2+\frac{1}{4\tau^2}
\label{ges7}
\end{equation}

\noindent and the energy (frequency) gap $\omega_g=\frac{1}{2\tau}$.

The solution to \eqref{ges2} can then be written in the form of space-decaying field as

\begin{equation}
u\propto\exp\left(-\frac{x}{2c\tau}\right)\exp\left(i(kx-\omega t)\right)
\label{ges5}
\end{equation}

We note that the first derivative in the telegraph equation \eqref{ges11} yielding GMS can be eliminated using a similar transformation: setting $u=\phi\exp\left(-\frac{t}{2\tau}\right)$ \cite{math2,math1,kosh} in (\ref{ges11}) gives the Klein-Gordon equation with the negative squared mass $-\frac{1}{4\tau^2}$:

\begin{equation}
c^2\frac{\partial^2\phi}{\partial x^2}=\frac{\partial^2\phi}{\partial t^2}-\frac{\phi}{4\tau^2}
\label{t1}
\end{equation}

\noindent and the same spectrum as in (\ref{gms5}).

Although Eqs. \eqref{ges11} and \eqref{t1} are related, they describe different physical systems and effects. Eq. \eqref{ges11} is based on physics of real systems such as liquids and describes dissipation in these systems with ensuing GMS. On the other hand, Eq. (\ref{t1}) is used to describe a hypothetical tachyon \cite{tachyons} or instability not present in equilibrium systems discussed here.

Similarly, although Eqs. \eqref{ges2} and \eqref{t2} are related, the physical difference is that we don't assume massive particles from the outset: our picture is based on the dissipation of plane waves generically related to the anharmonicity (coupling) of the interaction potential as in the case of GMS in Eq. \eqref{ges11}, with dissipation represented by an external force in Eq. \eqref{ges2}.

The gapped energy state (GES) can also be seen by considering the interaction potential between particles due to the scalar field $u$ in (\ref{ges2}). The propagator corresponding to Eq. (\ref{ges2}) is $\frac{1}{\omega^2-c^2k^2+\frac{ikc}{\tau}}$. Its spatial Fourier transform in one-dimensional example is

\begin{equation}
D(\omega,r)=\frac{1}{2\pi}\int\limits_{-\infty}^\infty\frac{e^{ikr}}{\omega^2-c^2k^2+\frac{ikc}{\tau}}dk
\label{inter1}
\end{equation}

Evaluating the integral gives

\begin{equation}
D(\omega,r)=e^{-\frac{r}{2c\tau}}\frac{\sin(k_rr)}{k_r}
\label{inter2}
\end{equation}

\noindent where $k_r=\frac{1}{c}\sqrt{\omega^2-\frac{1}{4\tau^2}}$.

$D(\omega,r)$ in \eqref{inter2} exponentially decreases with distance, in contrast to the photon propagator ($D(\omega,r)=-e^{i\omega r}/r$ in three-dimensional case \cite{landau}) and has the form of the correlator with the energy (mass) gap. Short-range interaction in Eq. \eqref{inter2} and the mass gap in Eq. \eqref{ges7} are related to the essence of Yukawa potential. The difference here is that we don't assume the massive particles from the outset as mentioned earlier.

%For $k_r=0$, corresponding to $\omega=\omega_g=\frac{1}{2\tau}$ in \eqref{ges6}, \eqref{inter2} reduces to $D(r)=re^{-\frac{r}{2c\tau}}$.

The same result for the energy gap can be found by seeking the solution of (\ref{ges2}) in the form $u=u_0\exp\left(i(kx-\omega t)\right)$ as before and solving the resulting quadratic equation for real $\omega$ and complex $k$ as in the case of the space-decaying field and skin effect \cite{math2,skin} (see discussion before Eq. \eqref{gms7}).

Eqs. (\ref{ges7})-(\ref{ges5}) is the main result of this paper. It shows that the gapped energy state (\ref{ges7}) can emerge as an accompanying result of dissipation of plane waves (\ref{ges5}) which, in turn, is generically related to the anharmonicity of the interaction potential (\ref{lzee}) as discussed earlier. This suggests that the spectra in Fig. 1 could be part of the same and more general mechanism generating both GMS and GES depending on the type of the external force.

If both types of external forces given by first derivatives in Eqs. \eqref{ges11} and \eqref{ges2} are present, the equation of motion can be written as

\begin{equation}
\frac{\partial^2u}{\partial t^2}=c^2\frac{\partial^2u}{\partial x^2}-\alpha_1\frac{1}{\tau}\frac{\partial u}{\partial t}+\alpha_2\frac{c}{\tau}\frac{\partial u}{\partial x}
\label{comb}
\end{equation}

\noindent where $\alpha_1$ and $\alpha_2$ are positive constants.

Eliminating both first derivatives using the standard transformation $u=\phi\exp\left(-\frac{\alpha_1t}{2\tau}\right)\exp\left(-\frac{\alpha_2x}{2c\tau}\right)$ \cite{math2,math1} gives

\begin{equation}
c^2\frac{\partial^2\phi}{\partial x^2}=\frac{\partial^2\phi}{\partial t^2}+(\alpha_2^2-\alpha_1^2)\frac{\phi}{4\tau^2}
\label{comb2}
\end{equation}

We observe that GMS and GES compete. $\alpha_1<\alpha_2$, $\alpha_1>\alpha_2$ and $\alpha_1=\alpha_2$ correspond to GES, GMS and gapless state, respectively. For $\alpha_1<\alpha_2$, the energy (mass) gap in \eqref{comb2} depends on the GMS contribution.

The properties of the external force $F_d\propto\frac{c}{\tau}\frac{\partial u}{\partial x}$ in Eq. \eqref{ges2} can now be discussed by comparing its sign to that of velocity $v=\frac{\partial u}{\partial t}$. Taking the real part of \eqref{ges5}, $F_d$ and $v$ can be written as

\begin{eqnarray}
\begin{split}
& F_d\propto\frac{c}{\tau}\frac{\partial u}{\partial x}=-\frac{c\omega}{\tau}e^{-\frac{x}{2c\tau}}\sin(kx-\omega t+\delta)\\
& \tan\delta=\frac{1}{2c\tau k}=\frac{\omega_g}{ck}\\
& v=v_0e^{-\frac{x}{2c\tau}}\sin(kx-\omega t)
\end{split}
\label{fandv}
\end{eqnarray}

\noindent where $v_0$ is the initial velocity and the angle $\delta$ defines the phase shift between $F_d$ and $v$.

For large $k$, $k\gg\frac{1}{2c\tau}$ ($\omega_g$ is close to 0), $\delta=0$, and $F_d\propto-v$, as is the case for the usual friction force. For an arbitrary $\delta$, $F_d$ and $v$ have opposite signs when $0<\psi<\pi-\delta$ and $\pi<\psi<2\pi-\delta$, where $\psi$ is $\psi=kx-\omega t$. This gives the range of $\psi$ of $2\pi-2\delta$, or $1-\frac{\delta}{\pi}$ in terms of the fraction of $2\pi$. Accordingly, $F_d$ and $v$ have the same sign during the fraction $\frac{\delta}{\pi}$. For $k$ corresponding to the propagation range $2c\tau$ in Eq. \eqref{ges5}, $\delta=\frac{\pi}{4}$ from Eq. \eqref{fandv}, and $F_d$ and $v$ have the opposite and same signs during fractions $\frac{3}{4}$ and $\frac{1}{4}$, respectively. These fractions become equal for $k\ll\frac{1}{2c\tau}$ and $\delta=\frac{\pi}{2}$, although waves with these $k$ are not propagating because their wavelengths are longer than the propagation range according to \eqref{fandv}. Therefore, Eq. \eqref{ges2} can be interpreted as a vibration of the rod induced at different $\omega$ and $k$ in a slowly moving medium, with the result that (a) the medium opposes the motion of the rod locally and at short wavelengths with the usual friction force $F_d\propto-v$, and (b) the medium alternates between opposing and speeding up the motion of the rod at longer wavelengths, with larger $\delta$ corresponding to the larger range of $k$ at which the medium speeds up the motion. According to Eq. \eqref{fandv}, the energy (mass) gap $\omega_g$ grows with $\delta$ and hence increases with the range of $k$ at which the medium speeds up the motion of the rod.

We make two further remarks regarding the energy gap. First, the GES is analogous to the mass gap whose emergence in strongly-interacting field theories has been of interest. In the Higgs mechanism, the gauge field acquires mass by interacting with the anharmonic Higgs field. In our picture, the energy gap can emerge as an accompanying effect of the dissipation due to the anharmonicity or self-interaction of the field itself. In interesting similarity to the GMS \eqref{gms5}, the Higgs potential also involves the negative mass squared.

Second, the dissipation of fields in Eqs. (\ref{gms4}) and (\ref{ges5}) is related to the sign of the dissipative terms with first derivatives in Eqs. (\ref{gms1}) or (\ref{ges11}) and (\ref{ges2}). Reversing the sign in these terms does not affect the DR with $k$-gap (\ref{gms5}) and energy gap (\ref{ges7}) but results in fields (\ref{gms4}) and (\ref{ges5}) increasing with time and distance, respectively. The nature of this increase is revealed in the Lagrangian formulation of dissipation necessitating two fields \cite{pre,lagr1}. The Lagrangian describing the GMS is

\begin{equation}
L=\frac{\partial\phi_1}{\partial t}\frac{\partial\phi_2}{\partial t}-c^2\frac{\partial\phi_1}{\partial x}\frac{\partial\phi_2}{\partial x}+\frac{1}{2\tau}\left(\phi_1\frac{\partial\phi_2}{\partial t}-\phi_2\frac{\partial\phi_1}{\partial t}\right)
\label{l1}
\end{equation}

\noindent resulting in $c^2\frac{\partial^2\phi_1}{\partial x^2}=\frac{\partial^2\phi_1}{\partial t^2}+\frac{1}{\tau}\frac{\partial\phi_1}{\partial t}$ and $c^2\frac{\partial^2\phi_2}{\partial x^2}=\frac{\partial^2\phi_2}{\partial t^2}-\frac{1}{\tau}\frac{\partial\phi_2}{\partial t}$, where the first equation is identical to (\ref{gms1}) and the second equation gives the solution growing in time. Both fields correspond to loss and gain systems \cite{bender}. The role of the gain system in the Lagrangian formulation is to absorb the energy of the dissipating (loss) system. The Hamiltonian of the composite system is finite and bound from below \cite{pre,lagr1}. We note that Schwinger-Keldysh approach is generally used to discuss dissipation in the quantum field theory (see, e.g., Ref. \cite{kamenev}). We have recently shown \cite{lagr1} that \eqref{l1} is consistent with the Schwinger-Keldysh approach.

The Lagrangian describing the GES can be written similarly to \eqref{l1}:

\begin{equation}
L=\frac{\partial\phi_1}{\partial t}\frac{\partial\phi_2}{\partial t}-c^2\frac{\partial\phi_1}{\partial x}\frac{\partial\phi_2}{\partial x}+\frac{c}{2\tau}\left(\phi_1\frac{\partial\phi_2}{\partial x}-\phi_2\frac{\partial\phi_1}{\partial x}\right)
\label{l2}
\end{equation}

\noindent and gives the same equation for $\phi_2$ as \eqref{ges2}. The equation for $\phi_1$ represents the gain system. The Hamiltonian of the composite system is finite and bound from below, as in the case of \eqref{l1}.

In summary, we have shown the energy gap can accompany the dissipation of fields due to anharmonicity and emerge explicitly in the dispersion relation. It would be interesting to understand what specific features of the anharmonic terms in \eqref{lzee} promote the gaps in energy and momentum and to extend the proposed theory to other fields including vector gauge fields.

I am grateful to M. Baggioli, V. V. Brazhkin, R. Russo, C. White and A. Zaccone for discussions and EPSRC for support.

\end{document}